\documentclass[aps,prc,twocolumn,floatfix]{revtex4-1}
\usepackage{amsmath}
\usepackage[english]{babel}
\usepackage{amssymb}
\usepackage{graphicx}
\usepackage{epstopdf}
\usepackage{natbib}
\usepackage{epsfig}
\usepackage{bm}
\usepackage{hyperref}
\usepackage{dcolumn}
\usepackage{color}

\def\tra{{\rm Tr}\,}

\begin{document}

\title{\sc\Large{Effects of strong magnetic fields on quark matter and $\pi^0$
properties within nonlocal chiral quark models}}

\author{D. G\'omez Dumm$^{1,2}$}
\author{M.F. Izzo Villafa\~ne$^{1,3}$}
\author{S. Noguera$^{4}$}
\author{V. Pagura$^4$}
\author{N.N. Scoccola$^{1,2,5}$}

\affiliation{$^{1}$ CONICET, Rivadavia 1917, (1033) Buenos Aires, Argentina}
\affiliation{$^{2}$ IFLP, CONICET $-$ Dpto.\ de F\'{\i}sica, Fac.\ de Cs.\
Exactas, Universidad Nacional de La Plata, C.C. 67, (1900) La Plata, Argentina}
\affiliation{$^{3}$ Physics Department, Comisi\'{o}n Nacional de Energ\'{\i}a At\'{o}mica, }
\affiliation{Av.\ Libertador 8250, (1429) Buenos Aires, Argentina}
\affiliation{$^{4}$ Departamento de F\'{\i}sica Te\'{o}rica and IFIC, Centro Mixto
Universidad de Valencia-CSIC, E-46100 Burjassot (Valencia), Spain}
\affiliation{$^{5}$ Universidad Favaloro, Sol{\'{\i}}s 453, (1078) Buenos Aires, Argentina}

\begin{abstract}
We study the behavior of strongly interacting matter under a strong external
magnetic field in the context of chiral quark models that include nonlocal
interactions. In particular, we analyze the influence of a constant magnetic
field on the chiral quark condensates at zero and finite temperature,
studying the deconfinement and chiral restoration critical temperatures and
discussing the observed ``magnetic catalysis'' and ``inverse magnetic
catalysis'' effects. In addition, we analyze in this framework the behavior
of the $\pi^0$ mass and decay constant. The predictions of nonlocal chiral
quark models are compared with results obtained in lattice QCD.
\end{abstract}


\maketitle

\renewcommand{\thefootnote}{\arabic{footnote}}
\setcounter{footnote}{0}

\section{Introduction}
\label{intro} The study of the behavior of strongly interacting matter under
intense external magnetic fields has gained significant interest in the last
years. The corresponding theoretical analyses require in general to deal
with quantum chromodynamics (QCD) in nonperturbative regimes, therefore most
studies are based either in the predictions of effective models or in the
results obtained from lattice QCD (LQCD) calculations. In fact, in view of
the theoretical difficulty, most works concentrate on the situations in
which one has a uniform and static external magnetic field.

In this work we study the features of QCD phase transitions and the
properties of the $\pi^0$ meson under an intense external magnetic field
$\vec B$ (for recent reviews on this subject see
e.g.~Refs.~\cite{Andersen:2014xxa,Miransky:2015ava}). At zero temperature,
both the results of low-energy effective models of QCD and LQCD calculations
indicate that light quark-antiquark condensates should behave as increasing
functions of $B$, which is usually known as ``magnetic catalysis''. On the
contrary, close to the chiral restoration temperature, LQCD calculations
carried out with realistic quark masses~\cite{Bali:2011qj,Bali:2012zg} show
that the condensates behave as nonmonotonic functions of $B$, and this leads
to a decrease of the transition temperature when the magnetic field is
increased. This effect is known as ``inverse magnetic catalysis'' (IMC). In
addition, LQCD calculations predict an entanglement between the chiral
restoration and deconfinement critical temperatures~\cite{Bali:2011qj}.
These findings have become a challenge to model calculations. Indeed, most
naive effective approaches to low energy QCD predict that the chiral
transition temperature should grow with $B$, i.e., they do not find IMC. In
this contribution we discuss this issue in the framework of nonlocal
chiral quark models~\cite{Pagura:2016pwr, GomezDumm:2017iex}. It is seen
that nonlocal models are able to describe, at the mean field level, not only
the IMC effect but also the entanglement between chiral restoration and
deconfinement transition temperatures. Moreover, within these models we
study the mass and decay constant of the $\pi^0$
meson~\cite{GomezDumm:2017jij}, showing that the behavior of these
quantities with the external magnetic field is also in agreement with LQCD
results. The models considered here are a sort of nonlocal extensions of the
Nambu$-$Jona-Lasinio (NJL) model that intend to provide a more realistic
effective approach to QCD. Actually, nonlocality arises naturally in the
context of successful descriptions of low-energy quark dynamics, and it has
been shown~\cite{Noguera:2008} that nonlocal models can lead to a momentum
dependence in quark propagators that is consistent with LQCD results.
Moreover, in this framework it is possible to obtain an adequate description
of the properties of light mesons at both zero and finite temperature (see
e.g.~\cite{Contrera:2009hk,Carlomagno:2013ona} and references therein).

The article is organized as follows. In Sect.~II we introduce the formalism
to deal with a nonlocal NJL-like model in the presence of the magnetic field
at zero temperature. Then we extend this formalism to a finite temperature
system, including the coupling to a background gauge field, and sketch the
analytical calculations required to study $\pi^0$ meson properties. In
Sect.~III we quote our numerical results, discussing the behavior of the
different relevant quantities as functions of the magnetic field and/or
temperature. Finally, in Sect.~IV we summarize our results and present our
conclusions.

\section{Theoretical formalism}
\label{sec-1}

Let us start by stating the Euclidean action for our nonlocal NJL-like
two-flavor quark model,
\begin{equation}
S_E = \int d^4 x \ \left\{ \bar \psi (x) \left(- i \rlap/\partial
+ m_c \right) \psi (x) -
\frac{G}{2} j_a(x) j_a(x) \right\} \ .
\label{action}
\end{equation}
where $m_c$ is the current quark mass (same for $u$
and $d$ quarks). The currents $j_a(x)$ are given by
\begin{equation}
j_a (x) = \int d^4 z \  {\cal G}(z) \
\bar \psi(x+\frac{z}{2}) \ \Gamma_a \ \psi(x-\frac{z}{2}) \ ,
\label{cuOGE}
\end{equation}
where $\Gamma_{a}=(\leavevmode\hbox{\small1\kern-3.8pt\normalsize1},i\gamma
_{5}\vec{\tau})$, and the function ${\cal G}(z)$ is a nonlocal form factor
that characterizes the effective interaction. We introduce now in the
effective action a coupling to an external electromagnetic gauge field
$\mathcal{A}_{\mu}$. For a local theory this can be done by performing the
replacement
\begin{equation}
\partial_{\mu}\ \rightarrow\ D_\mu\equiv\partial_{\mu}-i\,\hat Q
\mathcal{A}_{\mu}(x)\ ,
\label{covdev}
\end{equation}
where $\hat Q=\mbox{diag}(q_u,q_d)$, with $q_u=2e/3$, $q_d = -e/3$, is the
electromagnetic quark charge operator. In the case of the nonlocal model
under consideration, the inclusion of gauge interactions requires also a
change in the nonlocal currents in Eq.~(\ref{cuOGE}), namely
\begin{equation}
\psi(x-z/2) \ \rightarrow\ \mathcal{W}\left(  x,x-z/2\right)  \; \psi(x-z/2)\ ,
\label{transport}
\end{equation}
and the corresponding change for $\bar \psi(x+z/2)$~\cite{Noguera:2008}.
Here the function $\mathcal{W}(s,t)$ is defined by
\begin{equation}
\mathcal{W}(s,t)\ =\ \mathrm{P}\;\exp\left[ -\, i \int_{s}^{t}dr_{\mu}\, 
\hat Q\,\mathcal{A}_{\mu}(r)
\right]  \ ,
\label{intpath}%
\end{equation}
where $r$ runs over a path connecting $s$ with $t$. As it is usually done,
we take it to be a straight line.

To proceed we bosonize the fermionic theory, introducing scalar and
pseudoscalar fields $\sigma(x)$ and $\vec{\pi}(x)$ and integrating out the
fermion fields. The bosonized action can be written as
\begin{equation}
S_{\mathrm{bos}}=-\ln\det\mathcal{D}_{x,x'}+\frac{1}{2G}
\int d^{4}x\;
\Big[\sigma(x)\sigma(x)+ \vec{\pi}(x)\cdot\vec{\pi}(x)\Big]\ ,
\label{sbos}
\end{equation}
where
\begin{eqnarray}
\mathcal{D}_{x,x'}   &  = & \delta^{(4)}(x-x')\,\big(-i\,\rlap/\!D + m_{c} \big)\,
\mathcal{G}(x-x') \, \gamma_{0}\, \times \nonumber \\
& & \mathcal{W} (x,\bar x)\,
\gamma_{0} \big[\sigma(\bar x) + i\,\gamma_5\,\vec{\tau}\cdot\vec{\pi}(\bar x) \big]
\, \mathcal{W}(\bar x,x') \ ,
\label{dxx}%
\end{eqnarray}
with $\bar x = (x+x')/2$. We consider the case of a constant and homogenous
magnetic field orientated along the 3-axis, choosing the Landau gauge, in
which one has $\mathcal{A}_\mu = B\, x_1\, \delta_{\mu 2}$. In addition, we
assume that the field $\sigma$ has a nontrivial translational invariant mean
field value $\bar{\sigma}$, while the mean field values of pseudoscalar
fields $\pi_{i}$ are zero. In this way, within the mean field approximation
(MFA) we get
\begin{equation}
\mathcal{D}^{\mbox{\tiny MFA}}_{x,x'} \ = \ {\rm diag}
(\mathcal{D}^{\mbox{\tiny MFA},u}_{x,x'}\mathcal{D}^{\mbox{\tiny
MFA},d}_{x,x'})\ ,
\end{equation}
where
\begin{eqnarray}
\mathcal{D}^{\mbox{\tiny MFA},f}_{x,x'} & = &
\ \delta^{(4)}(x-x') \left( -i \rlap/\partial - q_f B
\, x_1 \gamma_2 + m_c \right) +
\nonumber \\
& & \hspace{-1.1cm} \bar\sigma \,\mathcal{G}(x-x') \,
\exp\left[i  (q_f B/2)\, (x_1+x'_1) \, (x_2-x'_2)\right]\, .
\label{df}
\end{eqnarray}
To deal with this operator it is convenient to introduce its Ritus transform
$\mathcal{D}^{\mbox{\tiny MFA},f}_{\bar p,\bar p\,'}$, defined by
\begin{equation}
\mathcal{D}^{\mbox{\tiny MFA},f}_{\bar p,\bar p\,'} = \int d^4x \ d^4x' \
\bar{\mathbb{E}}_{\bar p} (x)  \ \mathcal{D}^{\mbox{\tiny MFA},f}_{x,x'}  \
\mathbb{E}_{\bar p\,'} (x')\ ,
\label{dpp}
\end{equation}
where $\mathbb{E}_{\bar p} (x)$ and $\bar{\mathbb{E}}_{\bar p} (x)$, with
$\bar p=(k,p_2,p_3,p_4)$, are Ritus functions~\cite{Ritus:1978cj}. The index
$k$ is an integer that labels the Landau energy levels. Using the properties
of Ritus functions, after some calculation we
obtain~\cite{Pagura:2016pwr,GomezDumm:2017iex}
\begin{equation}
\mathcal{D}^{\mbox{\tiny MFA},f}_{\bar p,\bar p\,'} =
(2\pi)^4\delta_{kk'}\,\delta(p_2-
p_2^{\;\prime})\,\delta(p_3-p_3^{\;\prime})\, \delta(p_4-p_4^{\;\prime}) \ \mathcal{D}^f_{k, p_\parallel}\ ,
\label{diag}
\end{equation}
where
\begin{equation}
\mathcal{D}^f_{k, p_\parallel} \ = \ P_{k,{s_f}} \, \Big(\! -\! s_f\sqrt{2 k
|q_f B|}\; \gamma_2 +   p_\parallel \cdot \gamma_\parallel \Big) +
\sum_{\lambda=\pm} M^{\lambda,f}_{k,p_\parallel}\, \Delta^\lambda\ .
\label{twopoint}
\end{equation}
Here we have introduced the definitions $s_{f} = {\rm sign}(q_f B)$,
$p_\parallel = (p_3,p_4)$, $\gamma_\parallel = (\gamma_3,\gamma_4)$,
$\Delta^+=\mbox{diag}(1,0,1,0)$, $\Delta^-=\mbox{diag}(0,1,0,1)$ and
$P_{k,\pm 1}=(1-\delta_{k0})\,\mathcal{I}+\delta_{k0}\,\Delta^\pm$.
In addition, we denote
\begin{eqnarray}
M^{\lambda,f}_{k,p_\parallel} & = &
\frac{4\pi}{|q_fB|}\,(-1)^{k_\lambda}
\int \frac{d^2p_\perp}{(2\pi)^2}\
\left( m_c + \sigma\ g(p_\perp^2 + p_\parallel^2) \right) \,\times
\nonumber \\
& & \exp(-p_\perp^2/|q_fB|) \, L_{k_\lambda}(2p_\perp^2/|q_fB|)\ .
\label{mpk}
\end{eqnarray}
where we have used the definitions $k_\pm = k - 1/2 \pm s_f/2\,$ and $p_\perp
= (p_1,p_2)$, while $g(p^2)$ is the Fourier transform of $\mathcal{G}(x)$
and $L_m(x)$ are Laguerre polynomials, with the usual convention $L_{-1}(x)
=0$.

Using the fact that $\mathcal{D}^{\mbox{\tiny MFA},f}$ is diagonal
in Ritus space the corresponding contribution to the MFA action
can be readily calculated. We obtain
\begin{eqnarray}
\frac{S^{\mbox{\tiny MFA}}_{\mathrm{bos}}}{V^{(4)}} & = & \frac{ \bar
\sigma^2}{2 G} - N_c \sum_{f=u,d} \frac{  |q_f B|}{2 \pi} \int \frac{d^2p_\parallel}{(2\pi)^2}
\,
\nonumber \\
& & \bigg[ \ln\left(p_\parallel^2 + {M^{\,\lambda_{\!
f},f}_{0,p_\parallel}\,}^2\,\right)
+  \sum_{k=1}^\infty \ \ln \Delta^f_{k,p_\parallel}\bigg]\ ,
\label{smfa}
\end{eqnarray}
where $\lambda_f = +\, (-)$ for $s_f = +1\,(-1)$, and
$\Delta^f_{k,p_\parallel}$ is defined by
\begin{eqnarray}
\Delta^f_{k,p_\parallel} & = & \left( 2 k |q_f B| + p_\parallel^2 +
M^{+,f}_{k,p_\parallel}\, M^{-,f}_{k,p_\parallel} \right)^2\! +
\nonumber \\
& & p_\parallel^2
\left( M^{+,f}_{k,p_\parallel} - M^{-,f}_{k,p_\parallel} \right)^2\ .
\end{eqnarray}
Here it is seen that the functions $M^{\pm,f}_{k,p_\parallel}$ play the role
of constituent quark masses in the presence of the external magnetic field.

We extend now the analysis to a system at finite temperature. This is done
by using the standard Matsubara formalism. To account for confinement
effects, we also include the coupling of fermions to the Polyakov loop (PL),
assuming that quarks move on a constant color background field $\phi = i
g\,\delta_{\mu 0}\, G^\mu_a \lambda^a/2$, where $G^\mu_a$ are the SU(3)
color gauge fields. We work in the so-called Polyakov gauge, in which the
matrix $\phi$ is given a diagonal representation $\phi = \phi_3 \lambda_3 +
\phi_8 \lambda_8$, taking the traced Polyakov loop $\Phi=\frac{1}{3} {\rm
Tr}\, \exp( i \phi/T)$ as an order parameter of the
confinement/deconfinement transition. We also include in the Lagrangian a
Polyakov-loop potential ${\cal U}\,(\Phi, T)$, which accounts for effective
gauge field self-interactions. The resulting scheme is usually denoted as
nonlocal Polyakov-Nambu-Jona-Lasinio (nlPNJL)
model~\cite{Contrera:2007wu,Hell:2008cc}. For definiteness we will consider
here a polynomial PL potential of the form proposed e.g.\ in
Ref.~\cite{Ratti:2005jh}.

The grand canonical thermodynamic potential of the system under
the external magnetic field is found to be given by
\begin{eqnarray}
\Omega^{\mbox{\tiny MFA}}_{B,T} & = & \frac{ \bar
\sigma^2}{2 G} \ - \ T \sum_{n=-\infty}^{\infty} \sum_{c,f}\ \frac{  |q_f B|}{2 \pi} \int \frac{d p_3}{2\pi}
\ \bigg[ \ln\Big({p_\parallel}_{nc}^2 + \nonumber \\
& & \hspace{-0.2cm} {M^{\lambda_{\!
f},f}_{0,{p_\parallel}_{nc}}}^{\!2}\, \Big)
+
\sum_{k=1}^\infty \ \ln\left(
\Delta^f_{k,{p_\parallel}_{nc}}\right)\bigg] + \ {\cal U}(\Phi ,T)\ ,
\end{eqnarray}
where we have defined ${p_\parallel}_{nc} = (p_3\,,\,(2n+1)\pi T+\phi_c)$.
The sums over color and flavor indices run over $c=r,g,b$ and $f=u,d$,
respectively, while the color background fields are $\phi_r = - \phi_g =
\phi_3$, $\phi_b = 0$. As usual in nonlocal models, it is seen that
$\Omega^{\mbox{\tiny MFA}}$ turns out to be divergent, thus it has to be
regularized. We take here a usual prescription in which we subtract a free
contribution and add it in a regularized form. This ``free'' contribution is
in fact the potential obtained in absence of the strong current-current
coupling (i.e.\ setting $\bar \sigma = 0$), but keeping the interaction with
the magnetic field and the PL. One has
\begin{equation}
\Omega^{{\mbox{\tiny MFA}},\rm reg}_{B,T}\ = \ \Omega^{\mbox{\tiny
MFA}}_{B,T}\, -\, \Omega^{\rm free}_{B,T}\, +\, \Omega^{\rm free,reg}_{B,T}\
.
\end{equation}
The explicit form of $\Omega^{\rm free,reg}_{B,T}$ can be found in
Ref.~\cite{GomezDumm:2017iex}. The values of $\bar\sigma$ and $\Phi$ can be
obtained by minimization of $\Omega^{{\mbox{\tiny MFA}},\rm reg}_{B,T}$, and
the magnetic field dependent quark condensates $\langle \bar q_f q_f\rangle$
can be calculated by taking the derivatives of the thermodynamic potential
with respect to the corresponding current quark masses. To make contact with
LQCD results given in Ref.~\cite{Bali:2012zg} we define the quantities
\begin{equation}
\Sigma^f_{B,T} \ = \ -\frac{2\, m_c}{S^4} \left[ \langle \bar q_f q_f \rangle^{\rm reg}_{B,T}
 - \langle \bar q q\rangle^{\rm reg}_{0,0} \right] + 1 \ ,
\label{defi}
\end{equation}
where $S= (135 \times 86)^{1/2}$~MeV. We also introduce the definitions
$\Delta \Sigma^f_{B,T} = \Sigma^f_{B,T} - \Sigma^f_{0,T}$, $\bar
\Sigma_{B,T} = (\Sigma^u_{B,T}+\Sigma^d_{B,T})/2$ and $\Delta \bar
\Sigma_{B,T} = (\Delta \Sigma^u_{B,T}+\Delta \Sigma^d_{B,T})/2\,$, which
correspond to the subtracted normalized flavor condensate, the normalized
flavor average condensate and the subtracted normalized flavor average
condensate, respectively.

In addition, from the expansion of $\mathcal{D}_{x,x'}$ in powers of the
fluctuations of meson fields around their mean field values it is possible
to obtain the theoretical expressions for the $\pi^0$ meson mass (at $T=0$)
within our model. One has
\begin{eqnarray}
-\log\det\mathcal{D} & = &  - \tra\log{\mathcal{D}^{\mbox{\tiny MFA}}} \, - \,
\tra({\mathcal{D}^{\mbox{\tiny MFA}}}^{-1}\,\delta\mathcal{D}) \, + \nonumber \\
& & \frac{1}{2}\,\tra({\mathcal{D}^{\mbox{\tiny MFA}}}^{-1}\,\delta\mathcal{D})^2 \, + \, \dots
\label{expansion}
\end{eqnarray}
By writing the trace in momentum space one gets
\begin{eqnarray}
S_{\rm bos}\big|_{(\delta\pi_3)^2} & = &
\frac{1}{2}\,\tra({\mathcal{D}^{\mbox{\tiny MFA}}}^{-1}\,\delta\mathcal{D})^2\Big|_{(\delta\pi_3)^2}
+ \nonumber \\
& &  \frac{1}{2G}\int \frac{d^4t}{(2\pi)^4}\ \delta\pi_3(t)\,\delta\pi_3(-t)
\nonumber \\
& = & \frac{1}{2}\int \frac{d^4t}{(2\pi)^4} \Big[F(t_\perp^2,t_\parallel^2) +
\frac{1}{G}\Big]
\delta\pi_3(t)\,\delta\pi_3(-t)\ ,\nonumber\\
\label{sbospi}
\end{eqnarray}
where $F(t_\perp^2,t_\parallel^2)$ is a function that involves the external
field $B$. Its explicit form can be found in Ref.~\cite{GomezDumm:2017jij}.
Choosing the frame in which the $\pi^0$ meson is at rest, its mass can be
obtained as the solution of the equation
\begin{equation}
F(0,-m_{\pi^0}^2) +\frac{1}{G}\ = \ 0\ .
\label{pimass}
\end{equation}

On the other hand, in the absence of external fields, the $\pi^0$ decay
constant is defined through the matrix element of the axial current ${\cal
J}_{A3}^\mu$ between the vacuum and the physical pion state, taken at the
pion pole. One has
\begin{equation}
\langle 0|\,{\cal J}^\mu_{A3}(x)\,|\tilde \pi_3(t)\rangle \ = \
i\,e^{-i(t\cdot x)}\,f(t^2)\, t^\mu\ ,
\label{fpidef}
\end{equation}
where $\tilde \pi_3(t) = Z_{\pi^0}^{-1/2}\pi_3(t)$ is the renormalized field
associated with the $\pi^0$ meson state, with $t^2 = -m_{\pi^0}^2$. In our
framework it is possible to obtain an analytical expression for the form
factor $f(t^2)$ under a static uniform magnetic field, defining the $\pi^0$
decay constant $f_{\pi^0}(B)$ as the value of this form factor at $t^2 =
-m^2_{\pi^0}(B)$ (it should be noticed, however, that in the presence of the
magnetic field further Lorentz structures are allowed for the matrix element
in Eq.~(\ref{fpidef}), and there could exist other nonzero form factors).
The wave function renormalization factor $Z_{\pi^0}^{1/2}$ is given by the
residue of the pion propagator at $t^2 = -m_{\pi^0}^2$, namely
\begin{equation}
Z_{\pi^0}^{-1} \ = \
\frac{dF(0,t_\parallel^2)}{dt_\parallel^2}\bigg|_{t_\parallel^2 =
-m_{\pi^0}^2}\ .
\label{zpi}
\end{equation}

The matrix element in Eq.~(\ref{fpidef}) can be obtained by introducing a
coupling between the current ${\cal J}^\mu_{A3}$ and an auxiliary axial
gauge field $W_3^\mu$, and taking the corresponding functional derivative of
the effective action. In the same way as discussed at the beginning of this
section, gauge invariance requires the couplings to this auxiliary gauge
field to be introduced through the covariant derivative and the parallel
transport of the fermion fields, see Eqs.~(\ref{covdev}) and
(\ref{transport}).
Assuming that the mean field value of the $\pi_3$ field vanishes, the pion
decay constant can be obtained by expanding the bosonized action up to first
order in $W_{3\mu}$ and $\delta\pi_3$. Writing
\begin{equation}
S_{\rm bos}\big|_{W_3\,\delta\pi_3} \ = \ \int \frac{d^4t}{(2\pi)^4}
\, F_\mu(t)\; W_{3\mu}(t)\;\delta\pi_3(-t)\ ,
\end{equation}
one finds
\begin{equation}
f_{\pi^0} \ = \ f(-m_{\pi^0}^2) \ = \ i\,Z_{\pi^0}^{1/2}\,\frac{t_\mu
F_\mu(t)}{t^2}\bigg|_{t_\perp^2=0,t_\parallel^2 = -m_{\pi^0}^2}\ .
\label{fpi}
\end{equation}

To find the function $F_\mu(t)$ we consider once again the expansion in
Eq.~(\ref{expansion}). In addition, we expand $\delta{\cal D}$ in powers of
$\delta \pi_3$ and $W_3$,
\begin{equation}
\delta\mathcal{D} \ = \ \delta\mathcal{D}_{W} + \delta\mathcal{D}_\pi +
\delta\mathcal{D}_{W\pi} +\ \dots\ ,
\label{deltadexp}
\end{equation}
which leads to
\begin{eqnarray}
S_{\rm bos}\big|_{W_3\,\delta\pi_3} & = &
-\,\tra({\mathcal{D}^{\mbox{\tiny MFA}}}^{-1}\,\delta\mathcal{D}_{W\pi})\, + \nonumber\\
& & \,\tra({\mathcal{D}^{\mbox{\tiny MFA}}}^{-1}\,\delta\mathcal{D}_{W}{\mathcal{D}^{\mbox{\tiny MFA}}}^{-1}\,
\delta\mathcal{D}_\pi)\ .
\label{sbosfpi}
\end{eqnarray}
The explicit expression for $t\cdot F(t)|_{t_\perp = 0}\,$ is given in
Ref.~\cite{GomezDumm:2017jij}. It is worth mentioning that the chiral
Goldberger-Treiman and Gell-Mann-Oakes-Renner relations remain valid in our
model in the presence of the external magnetic
field~\cite{GomezDumm:2017jij}.

\section{Numerical results}

To obtain numerical predictions for the above defined quantities it is
necessary to specify the particular shape of the nonlocal form factor
$g(p^2)$. We will show here the results corresponding to the often-used
Gaussian form
\begin{equation}
g(p^2) \ = \ \exp(-p^2/\Lambda^2)\ .
\end{equation}
Notice that, owing to Lorentz invariance, we need to introduce in the form
factor an energy scale $\Lambda$, which acts as an effective momentum
cut-off and has to be taken as an additional parameter of the model. Thus,
the free parameters to be determined are $m_c$, $G$ and $\Lambda$. We have
considered different parameter sets, obtained by requiring that the model
leads to the empirical values of the pion mass and decay constant, as well
as some phenomenologically acceptable value of the quark condensate at $B=0$
and $T=0$. Here we take in particular $(-\langle \bar q q\rangle^{\rm
reg}_{0,0})^{1/3} = 220$, 230 and 240~MeV. The corresponding parameter sets
can be found e.g.\ in Ref.~\cite{GomezDumm:2017iex}.

\begin{figure}[hbt]
\includegraphics[width=0.45\textwidth]{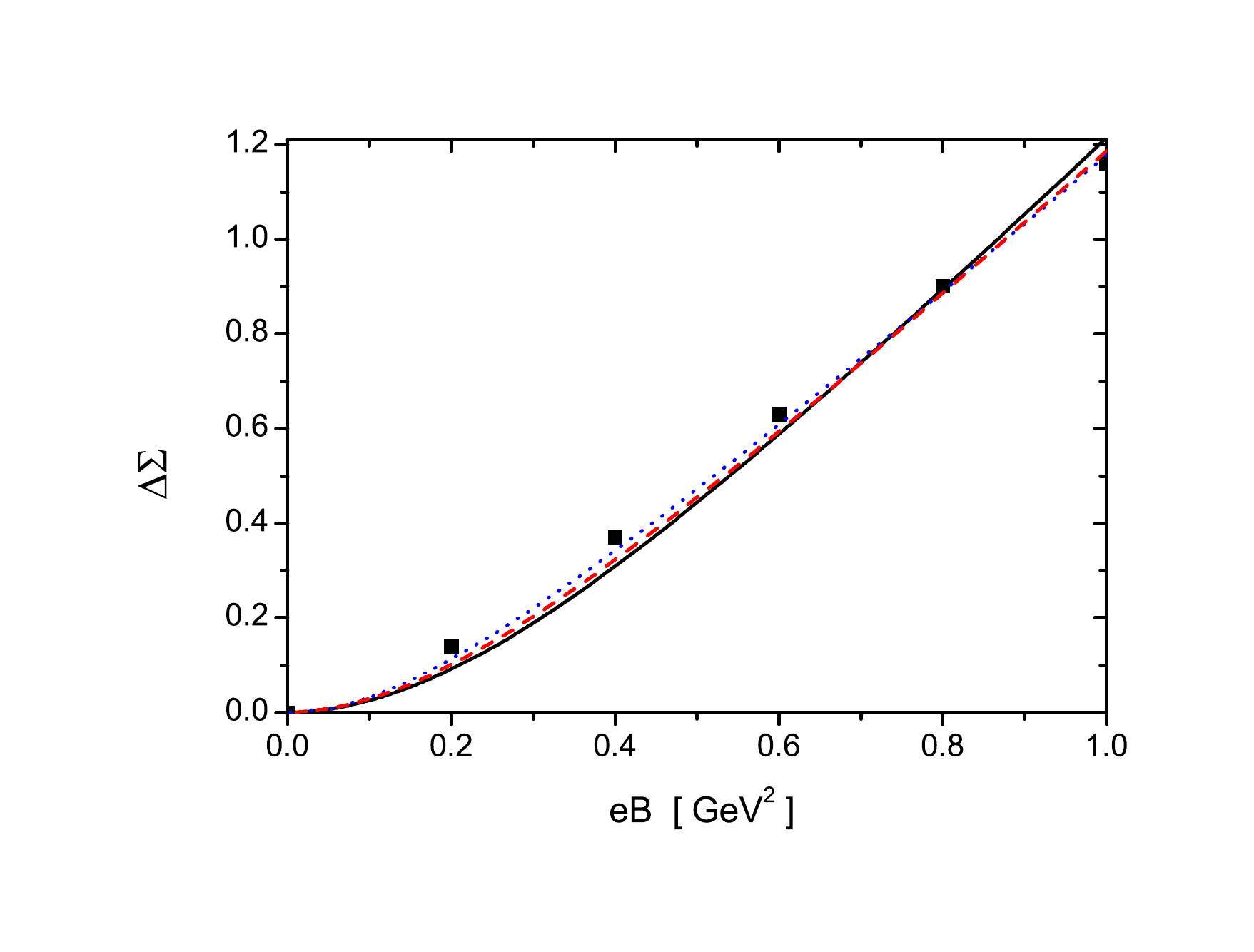}
\caption{Normalized condensates as functions of the magnetic field at $T =
0$. Solid (black), dashed (red) and dotted (blue) curves correspond to
parameterizations leading to $(- \langle\bar q q \rangle^{\rm
reg}_{0,0})^{1/3} = 220$, 230 and 240 MeV, respectively. Full square symbols
indicate LQCD results taken from Ref.~\cite{Bali:2012zg}.}
\label{fig1}
\end{figure}

Let us start by discussing our results for the condensates at zero
temperature. In Fig.~\ref{fig1} we show the predictions of our model for
$\Delta \bar \Sigma_{B,0}$ as functions of $eB$, together with LQCD data
from Ref.~\cite{Bali:2012zg}. Solid, dashed and dotted curves correspond to
$(-\langle \bar q_f q_f\rangle^{\rm reg}_{0,0})^{1/3} = 220$, 230 and 240
MeV, respectively. It can be seen that the predictions for $\Delta \bar
\Sigma_{B,0}$ are very similar for all parameter sets, and show
a very good agreement with LQCD results.

\begin{figure}[hbt]\begin{center}
\includegraphics[width=0.38\textwidth]{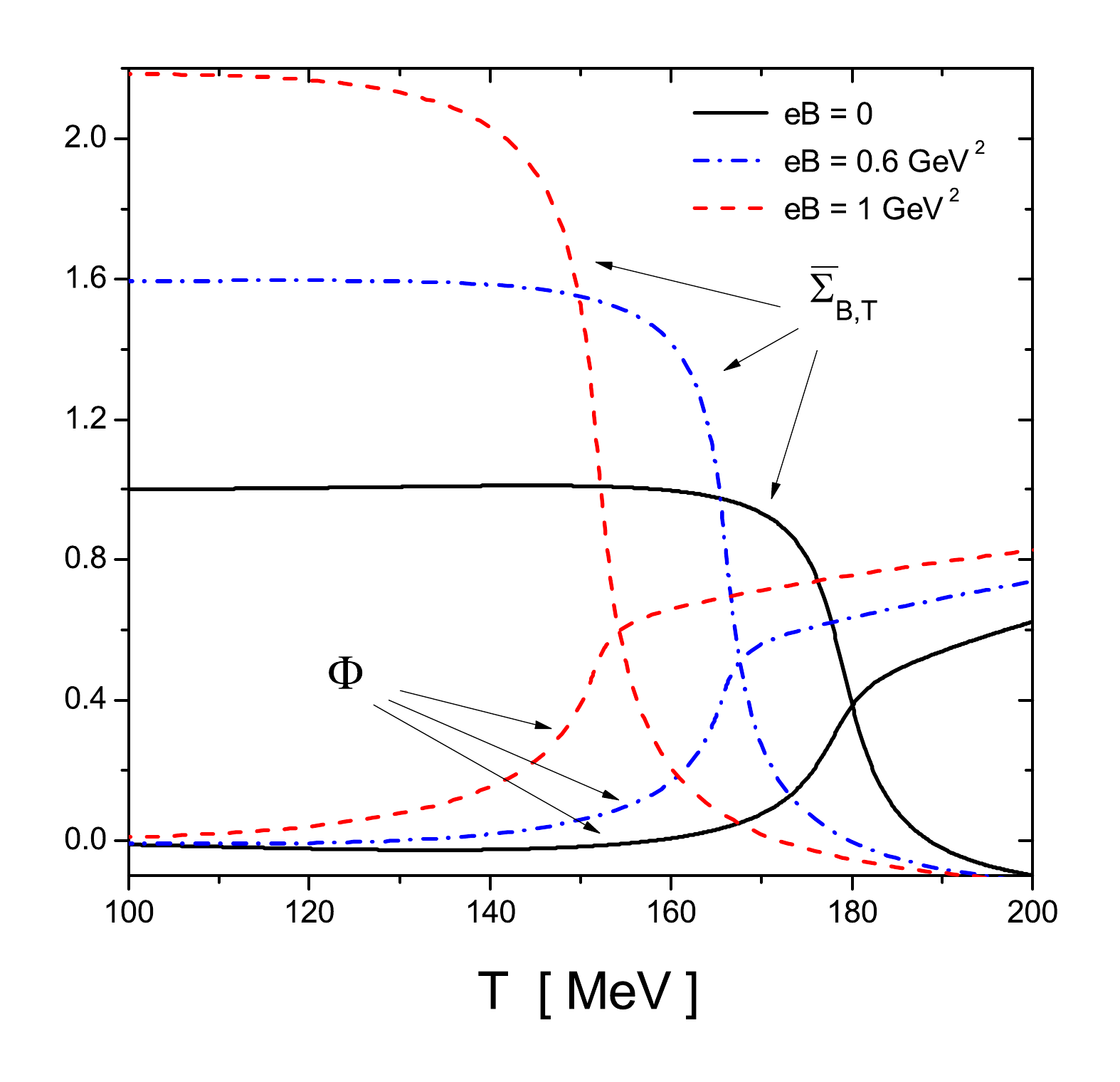}
\end{center}
\caption{Normalized flavor average condensate and traced
Polyakov loop as functions of the temperature, for three representative
values of $eB$.} \label{fig2}
\end{figure}

Next, we consider the results for a system at finite temperature. In
Fig.~\ref{fig2} we show the behavior of the averaged chiral condensate
$\bar\Sigma_{B,T}$ and the traced Polyakov loop $\Phi$ as functions of the
temperature, for three representative values of the external magnetic field
$B$, namely $B = 0$, 0.6 and 1 GeV$^2$. The curves correspond to parameter
sets leading to $(-\langle\bar q q \rangle^{\rm reg}_{0,0})^{1/3} = 230$
MeV. Given a value of $B$, it is seen from the figure that the chiral
restoration and deconfinement transitions proceed as smooth crossovers, at
approximately the same critical temperatures. For definiteness we take these
temperatures from the maxima of the chiral and PL susceptibilities, which we
define as the derivatives $\chi_{\rm ch} = -
\partial[(\langle\bar uu \rangle^{\rm reg}_{B,T} + \langle\bar dd
\rangle^{\rm reg}_{B,T})/2]/\partial T$ and $\chi_\Phi =
\partial\Phi/\partial T$, respectively.

\begin{table}[hbt]
\caption{Critical temperatures for $B=0$ and various parametrizations.}
\label{tab2}
\begin{center}
\begin{tabular}{l|ccc}
$(-\langle\, q \bar q\,\rangle_{0,0}^{\rm reg})^{1/3}$ (MeV) \ & \ 220 & 230 &
240 \ \\
Chiral $T_c$ (MeV) & \ \ 182.1 \ & \ 179.1 \ & \ 177.4 \ \ \\
Deconfinement $T_c$ (MeV)\ \ & \ \ 182.1 \ & \ 178.0 \ & \ 175.8 \ \
\end{tabular}
\end{center}
\end{table}

The chiral restoration and deconfinement critical temperatures obtained in
absence of external magnetic field are quoted in Table~\ref{tab2}. It is
seen that the splitting between both critical temperatures is below 5 MeV,
which is consistent with the results obtained in LQCD, and the values of
critical temperatures do not vary significantly with the parametrization. On
the other hand, the critical temperatures in Table~\ref{tab2} are found to
be somewhat higher than those obtained from LQCD, which lie around
160~MeV~\cite{Aoki:2009sc,Bazavov:2011nk}. It is worth mentioning that in
absence of the interaction with the Polyakov loop the values of $T_c$ drop
down to about 130~MeV~\cite{Pagura:2016pwr}.

\begin{figure}[hbt]
\begin{center}
\includegraphics[width=0.37\textwidth]{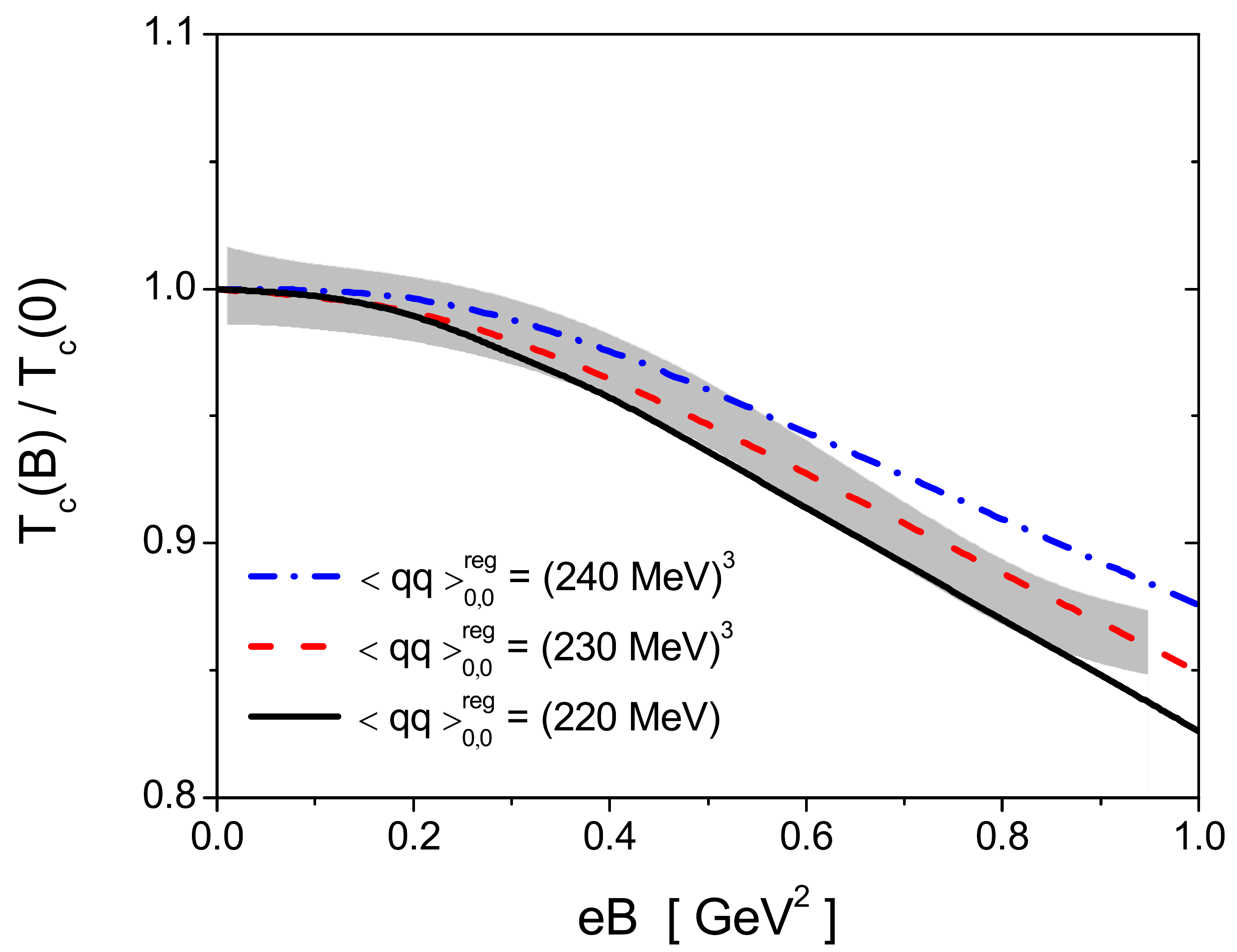}
\end{center}
\caption{Normalized critical temperatures as functions of $eB$. For
comparison, LQCD results of Ref.~\cite{Bali:2012zg} are indicated by the
gray band.}
\label{fig4}
\end{figure}

Let us discuss the effect of the magnetic field on the phase transition
features. From Fig.~\ref{fig2} it is seen that the splitting between the
chiral restoration and deconfinement critical temperatures remains very
small in the presence of the external field. In addition, the effect of
inverse magnetic catalysis is found. Indeed, contrary to what happens
e.g.~in the standard local NJL
model~\cite{Andersen:2014xxa,Miransky:2015ava}, in our model the chiral
restoration critical temperature becomes lower as the external magnetic
field is increased. This is related with the fact that the condensates do
not show in general a monotonic increase with $B$ for a fixed value of the
temperature. In Fig.~\ref{fig4} we plot our results for the chiral
restoration critical temperatures $T_c(B)$, normalized to the corresponding
values at vanishing external magnetic field. The gray band in indicates the
results obtained in LQCD, taken from Ref.~\cite{Bali:2012zg}. It is seen
that the critical temperatures decrease with $B$, i.e.~IMC is observed. As a
general conclusion, it can be stated that the behavior of the critical
temperatures with the external magnetic field is compatible with LQCD
results, for phenomenologically adequate values of the chiral condensate.

\begin{figure}[hbt]
\includegraphics[width=0.37\textwidth]{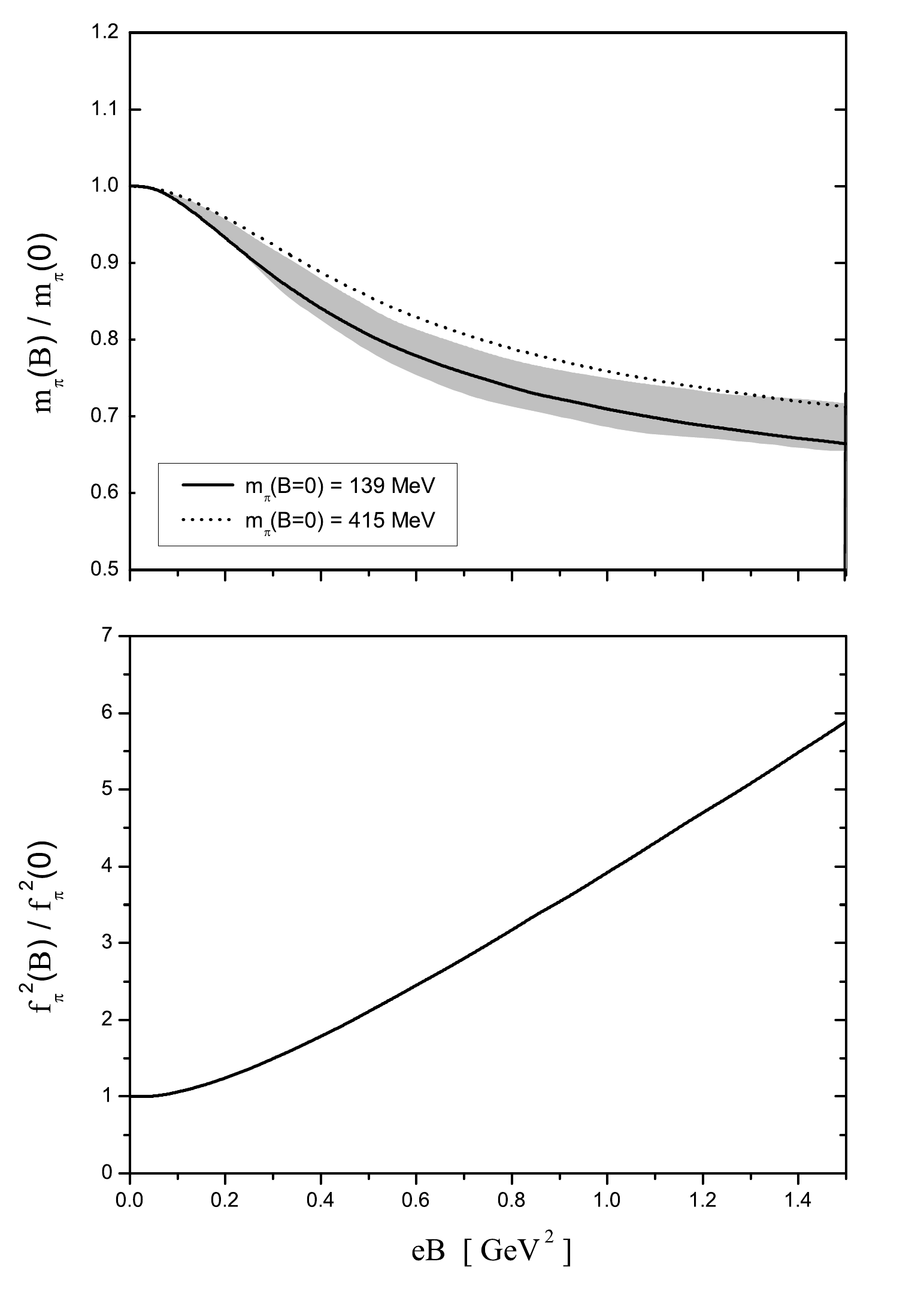}
\caption{Mass (upper panel) and decay constant (lower panel) of the $\pi^0$
meson as functions of $eB$, normalized to their values for $B=0$. In the
upper panel, the dotted line is obtained for a parameterization in which
$m_\pi = 415$~MeV, while the gray band corresponds to the results of lattice
QCD calculations quoted in Ref.~\cite{Bali:2017ian}.} \label{fig5}
\end{figure}

Finally, we study the behavior of the pion mass $m_{\pi^0}(B)$ and the
squared pion decay constant $f_{\pi^0}^{\,2}(B)$ for the above mentioned
parameter sets. Our results are shown in Fig.~\ref{fig5}, where once again
we consider the parameter set corresponding to $(-\langle\bar q q
\rangle^{\rm reg}_{0,0})^{1/3} = 230$~MeV. The curves have been normalized
to the $B=0$ empirical values of $m_{\pi^0}$ and $f_{\pi^0}^2$. As shown in
the upper panel of Fig.~\ref{fig5}, the $\pi^0$ mass is found to decrease
when $eB$ gets increased, reaching a value of about 65\% of $m_{\pi^0}(0)$
at $eB\simeq 1.5$ GeV$^2$, which corresponds to a magnetic field of about
$2.5\times 10^{20}$~G. We also include in the figure a gray band that
corresponds to recently quoted results from lattice QCD~\cite{Bali:2017ian}.
The latter have been obtained from a continuum extrapolation of lattice
spacing, considering a relatively large quark mass for which $m_\pi =
415$~MeV. For comparison, we also quote the results obtained within our
model by shifting $m_c$ to 56.3~MeV, which leads to this enhanced pion mass.
In general it is seen from the figure that our predictions turn out to be in
good agreement with LQCD calculations (notice that no extra adjustments have
been required). Concerning the pion decay constant, we find that it behaves
as an increasing function of $B$, which is consistent with the approximate
validity of the Gell-Mann-Oakes-Renner relation. It is also worth mentioning
that the curves in Fig.~\ref{fig5} are found to remain practically unchanged
when the value of the $B=0$ condensate used to fix the parameterization is
varied within the range from $-(220~{\rm MeV})^3$ to $-(250~{\rm MeV})^3$.

It can be seen that at large values of $eB$ our curves for both $T_c(B)$ and
$m_{\pi^0}(B)$ tend to fall more rapidly than the lattice bands. Notice,
however, that the validity of our approach is not guaranteed for too large
values of the magnetic field. In fact, this kind of quark models are in
general not trustable well above energy scales of about 1~GeV, where gluons
are expected to start showing up.

\section{Summary \& conclusions}

We have studied the behavior of strongly interacting matter under
a uniform static external magnetic field in the context of a
nonlocal chiral quark model. In this approach, which can be viewed
as an extension of the Polyakov-Nambu-Jona-Lasinio model, the
effective couplings between quark-antiquark currents include
nonlocal form factors that regularize ultraviolet divergences in
quark loop integrals and lead to a momentum-dependent effective
mass in quark propagators. We have worked out the formalism
introducing Ritus transforms of Dirac fields, which allows us to
obtain closed analytical expressions for the gap equations, the
chiral quark condensate and the quark propagator.

We have considered a Gaussian form factor, choosing some sets of model
parameters that allow to reproduce the empirical values of the pion mass and
decay constants. At zero temperature, for these parameterizations we have
determined the behavior of the subtracted flavor average condensate
$\Delta\bar\Sigma_{B,0}$ as a function of the external magnetic field $B$.
Our results show the expected magnetic catalysis (condensates behave as
growing functions of $B$), the curves being in quantitative agreement with
lattice QCD calculations with slight dependence on the parametrization.

We have also extended the calculations to finite temperature systems,
including the couplings of fermions to the Polyakov loop. We have defined
chiral and PL susceptibilities in order to study the chiral restoration and
deconfinement transitions, which turn out to proceed as smooth crossovers
for the considered polynomial PL potential. From our numerical calculations,
on one hand it is seen that, for all considered values of $B$, both
transitions take place at approximately the same temperature, in agreement
with LQCD predictions. On the other hand, it is found that for temperatures
close to the transition region $\Delta\bar\Sigma_{B,T}$ becomes a
nonmonotonic funtion of $B$, which eventually leads to the phenomenon of
inverse magnetic catalysis, i.e., a decrease of the critical temperature
when the magnetic field gets increased. This feature is also in qualitative
agreement with LQCD expectations. Moreover, in general we find a good
quantitative agreement with the results from LQCD calculations for the
behavior of the normalized critical temperatures with $B$. It is interesting
to compare the nonlocal models with approaches in which IMC is obtained by
considering some dependence of the effective couplings on $B$ and/or
$T$~\cite{Ayala:2014iba,Farias:2014eca}. The naturalness of the IMC behavior
in our framework can be understood by noticing that for a given Landau level
the associated nonlocal form factor turns out to be a function of the
external magnetic field, according to the convolution in Eq.~(\ref{mpk}).

Finally, we have studied the behavior of the $\pi^0$ mass and decay constant
with the magnetic field. Both quantities are found to show a very mild
dependence on the parametrization. It is worth noticing that our results for
the pion mass turn out to be in good agreement with available lattice QCD
calculations, with no need of extra ad-hoc assumptions.

\section*{Acknowledgements}

This work has been supported in part by CONICET and ANPCyT (Argentina),
grants PIP14-492, PIP12-449, and PICT14-03-0492, by UNLP (Argentina), Project X824, by the Mineco
(Spain), contract FPA2013-47443-C2-1-P, FPA2016-77177-C2-1-P, by Centro de Excelencia
Severo Ochoa Programme, grant SEV-2014-0398, and by
Generalitat Valenciana (Spain), grant PrometeoII/2014/066.

\end{document}